\documentclass{PoS}
\usepackage{amsmath,amssymb}

\title{Chiral symmetry breaking, instantons, and monopoles}

\ShortTitle{Chiral symmetry breaking, instantons, and monopoles}

\author{Adriano Di Giacomo\\
        University of Pisa, Department of Physics and INFN, Sezione di Pisa, Largo B. Pontecorvo 3, 56127, Pisa, Italy\\
        E-mail: \email{digiaco@df.unipi.it}}

\author{\speaker{Masayasu Hasegawa}\\
        Joint Institute for Nuclear Research, Bogoliubov Laboratory of Theoretical Physics, Dubna, 141980, Moscow, Russia\\
        E-mail: \email{hasegawa@theor.jinr.ru}}

\abstract{The purpose of this study is to show that monopoles induce the chiral symmetry breaking. In order to indicate the evidence, we add one pair of monopoles with magnetic charges to the quenched SU(3) configurations by a monopole creation operator, and investigate the propaties of the chiral symmetry breaking using the Overlap fermion. We show that instantons are created by the monopoles. The pseudoscalar meson mass and decay constant are computed from the correlation functions, and the renormalization constant $Z_{S}$ is determined by the non perturbative method. The renormalization group invariant chiral condensate in $\overline{\mbox{MS}}$-scheme at 2 [GeV] is evaluated by the Gell-Mann-Oakes-Renner formula, and the random matrix theory. Finally, we estimate the renormalization group invariant quark masses $\bar{m} = (m_{u} + m_{d})/2$, and $m_{s}$ in $\overline{\mbox{MS}}$-scheme at 2 [GeV]. The preliminary results indicate that the chiral condensate decreases and the quark masses become slightly heavy by increasing the number of monopole charges.}

\FullConference{The 33rd International Symposium on Lattice Field Theory\\
		14 -18 July 2015\\
		Kobe International Conference Center, Kobe, Japan}

\begin{document}

\section{Introduction}
The quark confinement is caused by the monopole condensation in QCD, that is beautifully explained from the dual superconductivity by t'Hoot and Mandeltams~\cite{tHooft1}. The instanton configurations induce the chiral symmetry breaking~\cite{Shuryak1,Diakonov1}. In this study we want to show that monopoles are related with the instantons and the chiral symmetry breaking, by using the Overlap fermion which preserves the chiral symmetry in the lattice gauge theory~\cite{Gins1}. In order to show the relation, we add monopoles with magnetic charges to SU(3) quenched configurations by the monopole creation operator~\cite{DiGiacomo1}. We compute the low-lying eigenvalues and eigenvectors from the gauge links of normal configurations and configurations with additional monopoles, by solving eigenvalue problems of the Overlap Dirac operator~\cite{DeGrand1,Giusti1}.

In previous work we have shown that the topological susceptibility in the continuum limit is properly computed, and the instanton density is consistent with the values computed from instanton liquid model in Ref.~\cite{Shuryak1}. The additional monopoles form long loops in QCD vacuum. Instantons are created by the additional monopoles, moreover, the additional monopoles do not affect detection of the fermion zero modes~\cite{DiGH1}. Our study in the Maximal Abelian gauge shows that the number of observed zero modes increases with the square root of the total physical length of monopole loops, and the number of instantons is in direct proportion to the total physical length of monopole loops~\cite{DiGH2}. Moreover, in the study of the random matrix theory, we have precisely computed the chiral condensate which is the order parameter of the chiral symmetry breaking. We have confirmed that the additional monopoles do not affect the low-lying eigenvalues of the Overlap Dirac operator, and found that the chiral condensate decreases by increasing the magnetic charges of the additional monopoles~\cite{DiGHP1}. This is an evidence that monopoles directly induce the chiral symmetry breaking.
 
However, those studies have been done by adding monopoles with charges to the small lattice volume ($V = 14^{4}$) and one lattice spacing ($\beta = 6.00$). Therefore, we have to investigate the values at the continuum limit and the finite lattice volume effects. In present study we add monopoles with magnetic charges to configurations of the larger lattices volumes and different parameters of lattice spacing. 

In this report, we count the number of instantons from the average squares of the topological charges, and compute the instanton density from the normal configurations. We show that one instantons of $\pm$ charge is created by one pair of monopoles with magnetic charges $m_{c} = \pm1$ [Section~\ref{sec:ins_mon_1}]. Next, we compute the pseudoscalar meson mass and decay constant from the correlation functions, and derive the renormalization constant $Z_{S}$ by the non perturbative method. The renormalization group invariant chiral condensate in the $\overline{\mbox{MS}}$-scheme at 2 [GeV] is evaluated by the Gell-Mann-Oakes-Renner (GMOR) formula, and the random matrix theory (RMT). The results of the chiral condensate which are computed from normal configurations are $\langle \bar{\psi}\psi\rangle_{\overline{MS}}^{GMOR} \ (2 \ \mbox{GeV}) = -0.0271(13)$ [GeV$^{3}$], $\langle \bar{\psi}\psi\rangle_{\overline{MS}}^{RMT} \ (2 \ \mbox{GeV}) = -0.0269(14)$ [GeV$^{3}$], $(r_{0} = 0.5 [\mbox{fm}])$. The renormalization group invariant quark masses, ${\bar{m} = (m_{u} + m_{d})/2}$ and $m_{s}$, in the $\overline{\mbox{MS}}$-scheme at 2 [GeV] are estimated. The results which are computed for normal configurations are $\bar{m}^{\overline{MS}} \ (2 \ \mbox{GeV}) = 4.0(4)$ [MeV], and $m_{s}^{\overline{MS}} \ (2 \ \mbox{GeV}) = 98 (9)$ [MeV], ($a^{-1} = 2.00(8)$ [GeV]).

 Lastly, the preliminary results show that the chiral condensate decreases with increasing the number of monopole charges. The quark masses become slightly heavy with increasing the number of monopole charges [Section~\ref{sec:chiral_mono_1}].

\section{Instantons and monopoles}\label{sec:ins_mon_1}
\subsection{Overlap fermions}

The Overlap fermion preserves the chiral symmetry in the lattice gauge theory, and has the fermion zero modes~\cite{Gins1}. The massless Overlap Dirac operator is computed from the massless Wilson Dirac operator, which is constructed from the gauge links of the configurations~\cite{DeGrand1,Giusti1}. In this study the (negative) mass parameter is $\rho = 1.4$, and the numbers of low-lying eigenvalues and eigenvectors of the Overlap Dirac operator are $\mathcal{O}(40 - 100)$, which are computed by solving the eigenvalue problems using the subroutines (ARPACK). In this section, the lattice spacing is computed from the interpolating function~\cite{Necco1}. The scale is $r_{0}$ = 0.5 [fm].

Let $n_{+}$ be the number of zero modes of chirality plus, $n_{-}$ the number of those with chirality minus. Those numbers are counted from exact zero modes in spectra of the Overlap Dirac operator. The topological charge is defined as $Q = n_{+} - n_{-}$. The topological susceptibility is computed from $\langle Q^{2} \rangle/V$. 

However, in our simulation, we never observed the zero modes of chiralities $n_{+}$ and $n_{-}$ in one configuration at the same time. Therefore, to check the reason, we compare the distribution of the topological charges with the results produced by the other group. Moreover, we compute the topological susceptibility at the continuum limit, by interpolating five data points ($\beta$ = 5.8124, 5.9044, 5.9890, 6.0000, and 6.0680) possessing the same physical volume $V/r_{0}^{4} = 49.96$ to the continuum limit. The fitting function is $\langle Q^{2} \rangle r_{0}^{4}/V = c_{1}\cdot (a/r_{0})^{2} + c_{0}$. The topological susceptibility at the continuum limit in this study is 
\begin{equation}
\chi = (194(3) \ [\mbox{MeV}])^{4}.
\end{equation}
The theoretical prediction is $\chi = (1.80\times 10^{2} \ \ [\mbox{MeV}])^{4}$~\cite{Witten_Veneziano1}. The distribution of the topological charges and the value of the topological susceptibility at the continuum limit are consistent with the results in Ref.~\cite{Debbio1}. Therefore, we can properly compute the topological charges. In our computation, the number of observed zero modes is exactly the same as the topological charge. Accordingly, the observed number of zero modes is the \textbf{net} number $n_{+} - n_{-}$.

\subsection{Instanton density}

The number of instantons $N_{i}$ is computed from the average square of the topological charges $\langle Q^{2}\rangle$. The relation $N_{i} = \langle Q^{2}\rangle$ is analytically derived~\cite{DiGH1}. Here, we evaluate the instantons density $\rho_{i}$ from the slope $A$ by fitting a linear function $ \langle Q^{2}\rangle = A\cdot V/r_{0}^{4} + B$ to 21 data points. The results are
\begin{equation}
 \rho_{i} =  8.0(2) \times 10^{-4} \ [\mbox{GeV}^{4}], \ B = -0.02(0.10), \ \chi^{2}/d.o.f. = 24.4/19.0.
\end{equation}
The intercept $B$ is zero, and $\chi^{2}/d.o.f.$ is 1.28. The number of instantons is in direct proportion to the physical volume. Our result is exactly consistent with the instanton liquid model $\rho_{i} =  8 \times 10^{-4} \ [\mbox{GeV}^{4}]$~\cite{Shuryak1}.

\subsection{Instantons and monopoles}
\begin{table*}[htbp]
  \small
  \caption{The simulation parameters and the fitted results. $A_{Pred.}$ and $B_{Pred.}$ are computed from our prediction. The fitting range $FR$ is $m_{c}$ unit.}\label{tb:Add_mon_fit_res_1}
{\renewcommand{\arraystretch}{1.0}
  \begin{tabular*}{\textwidth}{c @{\extracolsep{\fill}}ccccccccc} \hline\hline
$\beta$ \ \ \ \ & $a/r_{0}$ & $V$      & $V/r_{0}^{4}$ &$A_{Pred.}$ & $B_{Pred.}$ & $A$ & $B$ & $FR$ & $\chi^{2}/d.o.f.$ \\ \hline
5.9256  \ \ \ \ & 0.2129 & $14^{3}28$ & 157.89 &1.0(2)  & 11.3(6)& 1.1(2)  & 11.6(5) & 0 - 4 & 0.8/3.0  \\
6.0000  \ \ \ \ & 0.1863 & $14^{4}$   & 46.276 &1.00(3) & 3.19(7)& 1.07(5) & 3.07(10)& 0 - 4 & 8.5/3.0  \\
        \ \ \ \ & & $14^{3}28$ & 92.553 &1.00(14)& 6.6(3) & 0.94(16)& 5.9(3)  & 0 - 4 & 43.3/3.0 \\
        \ \ \ \ & & $16^{3}32$ & 157.89 &1.00(14)& 9.1(4) & 1.16(17)& 10.1(4) & 0 - 5 & 8.5/4.0  \\
6.0522  \ \ \ \ & 0.1705 & $18^{3}32$ & 157.89 &1.00(11)& 9.4(4) & 1.22(14)& 10.6(4) & 0 - 6 & 4.5/5.0  \\ \hline\hline
\end{tabular*}
}
\end{table*}
\begin{figure}[htbp]
  \begin{center}
    \includegraphics[width=150mm]{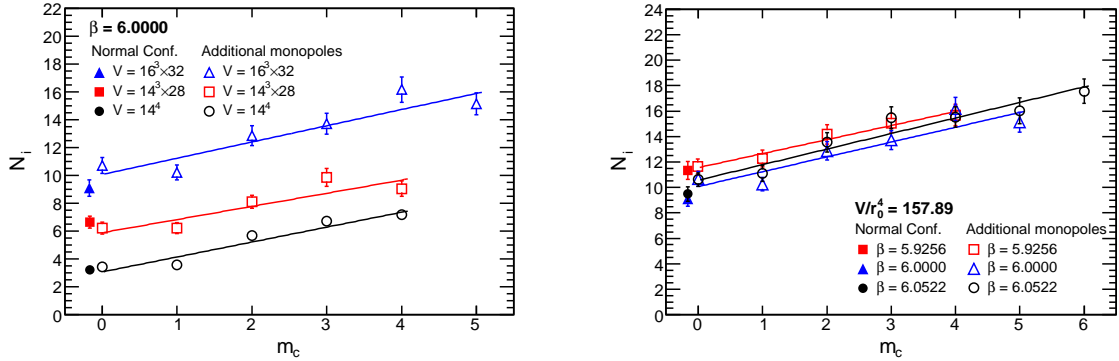}
  \end{center}
  \caption{The number of instantons $N_{i}$ versus the number of monopole charges $m_{c}$. The lattice spacing is fixed at $\beta = 6.0000$ (left figure). The physical lattice volume is fixed at $V/r_{0}^{4} = 157.89$ (right figure).}\label{fig:num_ins_add_mon_1}
\end{figure}
We confirm that we can properly count the number of instantons. We add one pair of a monopole and an anti-monopole varying magnetic charges $m_{c}$ to configurations, by the monopole creation operator~\cite{DiGiacomo1}. The monopole has positive magnetic charges $m_{c} = +1, +2, \cdots, +6$, and the anti-monopole has negative magnetic charges $m_{c} = -1, -2, \cdots, -6$. The total of the magnetic charges in the configuration set to zero. Henceforth, $m_{c}$ indicates that both charges $\pm m_{c}$ are added. We put the monopole and the anti-monopole holding the physical distance $D \approx 1.1$ [fm]. We generate configurations setting the lattice spacing at $\beta = 6.0000$ to check the finite lattice volume effects, and setting the physical lattice volume at $V/r_{0}^{4} = 157.89$ to check the continuum limit. The numbers of configurations are $\mathcal{O}(5\times10^{2}) \sim \mathcal{O}(7\times10^{2})$ for each parameter. The simulation parameters are in Table~\ref{tb:Add_mon_fit_res_1}. We fit a linear function $N_{i} = A\cdot m_{c} + B$ as show in Fig.~\ref{fig:num_ins_add_mon_1}, and compare the slope $A$ and the intercept $B$ with our prediction in Table~\ref{tb:Add_mon_fit_res_1}. The details of our prediction is explained in Ref.~\cite{DiGH1}. The slope $A$ and the intercept $B$ are consistent with our prediction. The finite lattice volume and the discretization do not affect the results. Therefore, one pair of monopoles with one positive charge and one negative charge creates one instanton of one positive or negative charge.

\section{Chiral symmetry breaking and monopoles}~\label{sec:chiral_mono_1}
We compute the renormalization constant $Z_{S}$, and evaluate the renormalization group invariant chiral condensate and quark masses in the $\overline{\mbox{MS}}$-scheme at 2 [GeV], using one hundred pairs of low-lying eigenvalues and eigenvectors. The lattice for this study is $V = 16^{3}\times32$, $\beta = 6.0000$. The number of the monopole charges $m_{c}$ is from zero to five, and the number of configurations $N_{conf}$ is 500 for each measurement.

\subsection{The pseudoscalar correlation, renormalization constant, and chiral condensate} 

In this study the pseudoscalar meson mass $m_{PS}$ is computed from the difference between the pseudoscalar correlation function $C_{PP}$ and the scalar correlation function $C_{SS}$, $C_{PP-SS} \equiv C_{PP} - C_{SS}$. The correlation functions of the pseudoscalar $C_{PP}$ and the scalar $C_{SS}$ are computed as in Ref.~\cite{DeGrand2}. In order to evaluate precisely the square of pseudoscalar meson mass at the chiral limit, we compute the correlation function $C_{PP-SS}$ at 30 different valence quark masses. The range of the valence quark masses is 0.00472 $ \leq am_{q} \leq $ 0.04721 (10 $ \leq m_{q} \leq $ 100 [MeV]). We determine the the pseudoscalar meson mass $m_{PS}$ and $G_{PP-SS}$ by fitting a following function~\cite{Gimenez1,Giusti5}, 
\begin{equation}
 \scalebox{0.9}{$\displaystyle  C_{PP-SS}(t) = \frac{G_{PP-SS}}{am_{PS}}\exp\left(-\frac{am_{PS}}{2}T\right)\cosh\left\{am_{PS}\left(\frac{T}{2} - t\right)\right\}. $}
\end{equation}
We then compute $Z_{S}$ as in Refs.~\cite{DiGHP1,Hernandez1}. The results are in Table~\ref{tb:condensate_quarkmass_1}. Next, we compute the chiral condensate from Gell-Mann-Oakes-Renner mass formula defined as in Ref.~\cite{GMOR0},
\begin{equation}
   \scalebox{0.9}{$\displaystyle a^{3}\langle\bar{\psi}\psi\rangle = - \lim_{am_{q} \rightarrow 0}\frac{(af_{P})^{2}(am_{PS})^{2}}{2(am_{q})}, \ \ af_{P} = 2(am_{q})\frac{\sqrt{G_{PP-SS}}}{(am_{PS})^{2}}. $}\label{eq:gmor_1}
\end{equation}
\begin{figure}[htbp]
  \begin{center}
    \includegraphics[width=150mm]{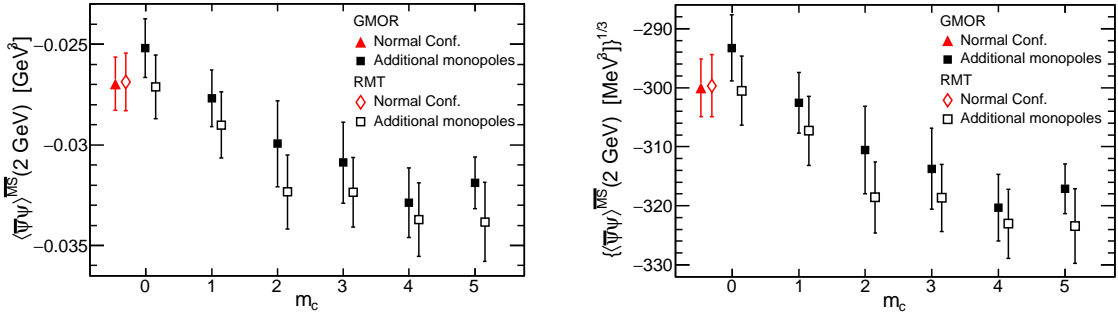}
  \end{center}
  \caption{The comparisons of the chiral condensates which are computed from the GMOR formula and the random matrix theory. The left figure is in [GeV$^{3}$] unit, and the right figure is in [MeV] unit.}\label{fig:Chiral_add_or_conti_mev_1}
\end{figure}
We convert $\langle\bar{\psi}\psi\rangle$ to the $\overline{\mbox{MS}}$-scheme at 2 [GeV] as follows: $\langle\bar{\psi}\psi\rangle^{\overline{MS}} (2 \ \mbox{GeV}) = \langle\bar{\psi}\psi\rangle Z_{S}/z, \ \ z = 0.72076$. The results comparing with the results produced from the random matrix theory are in Table~\ref{tb:condensate_quarkmass_1}. The scale is $r_{0}$ = 0.5 [fm]. The decay constant at the chiral limit that is computed from normal configurations is $f_{\pi} = 122.6(1.2)$ [MeV] ($r_{0}$ = 0.5 [fm])~\cite{Giusti5}. The chiral condensates that are computed from GMOR formula and RMT are consistent with each other. The chiral condensate decreases by increasing the monopole charges as shown in Fig~\ref{fig:Chiral_add_or_conti_mev_1}. 

\subsection{Quark mass}
\begin{table*}[htbp]
  \small
  \caption{The chiral condensates and the quark masses in the $\overline{\mbox{MS}}$ scheme at 2 [GeV]. $\langle\bar{\psi}\psi \rangle^{RMT}$ is computed from the scale parameter $\Sigma$ in RMT of the topological charge sector $|Q| = 1$. $\bar{m} = (m_{u} + m_{d})/2$. }\label{tb:condensate_quarkmass_1}
          {\renewcommand{\arraystretch}{1.0}
            \begin{tabular*}{\textwidth}{l @{\extracolsep{\fill}}ccccccc} \hline\hline
              $m_{c}$ & $Z_{S}$ & $\langle\bar{\psi}\psi\rangle_{\overline{MS}}^{GMOR}$ & $\langle\bar{\psi}\psi\rangle_{\overline{MS}}^{RMT}$ & $\{\langle\bar{\psi}\psi\rangle_{\overline{MS}}^{GMOR}\}^{1/3}$ & $\{\langle\bar{\psi}\psi\rangle_{\overline{MS}}^{RMT}\}^{1/3}$ & $\bar{m}^{\overline{MS}}$ & ${m_{s}}^{\overline{MS}}$\\
              & &  [GeV$^{3}$] & [GeV$^{3}$]& [MeV] & [MeV] & [MeV] & [MeV]   \\ \hline
N. C. \ \ \ & 0.99(3) & -0.0271(13) &  -0.0269(14) &  -300(5) & -300(5) &  4.0(4)&  98(9)  \\
 \ 0     & 1.00(4) & -0.0252(15) &  -0.0272(16) &  -293(6) & -301(6) &  3.9(4)&  94(9)  \\
 \ 1     & 1.00(3) & -0.0277(14) &  -0.0290(17) &  -303(5) & -307(6) &  3.9(6)&  96(12) \\
 \ 2     & 0.98(3) & -0.030(2)   &  -0.0323(18) &  -311(7) & -319(6) &  4.2(5)&  104(10)\\
 \ 3     & 0.97(3) & -0.031(2)   &  -0.0324(17) &  -314(7) & -319(6) &  4.5(5)&  110(11)\\
 \ 4     & 0.98(3) & -0.0329(17) &  -0.0337(18) &  -320(6) & -323(6) &  4.6(5)&  113(11)\\
 \ 5     & 0.99(3) & -0.0319(13) &  -0.034(2)   &  -317(4) & -323(6) &  4.8(5)&  116(11)\\\hline\hline
\end{tabular*}
}
\end{table*}
\begin{figure}[htbp]
  \begin{center}
    \includegraphics[width=150mm]{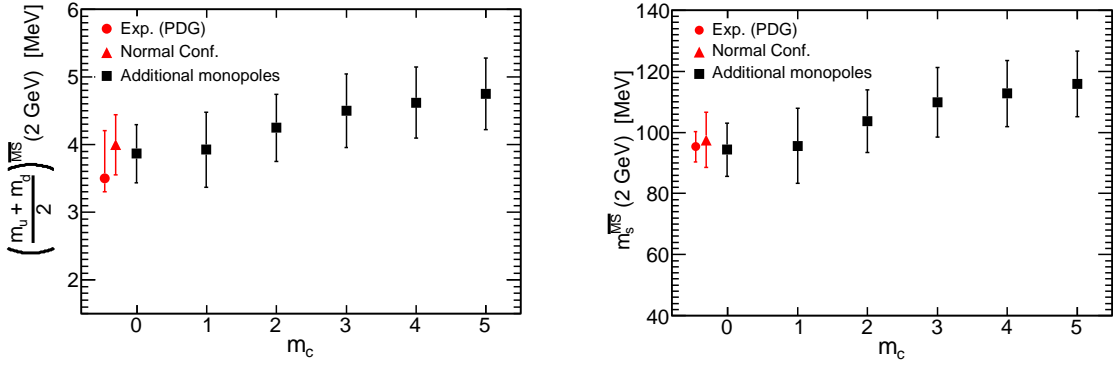}
  \end{center}
 \caption{The comparisons of the quark masses, $(m_{u} + m_{d})/2$ (left figure) and $m_{s}$ (right figure) in $\overline{\mbox{MS}}$-scheme at 2 [GeV]. The experimental results are $\bar{m} = 3.5_{-0.2}^{+0.7}$ [MeV], and $m_{s} = 95 (5)$ [MeV] (PDG).}\label{fig:quark_mass_1}
\end{figure}
Lastly, we estimate the quark masses, $\bar{m} = \frac{m_{u} + m_{d}}{2}$ and $m_{s}$, based on Ref.~\cite{Giusti5}. The mass ratio is known from the chiral perturbation theory as follows: $R = \frac{m_{s}}{\bar{m}} = 24.4 \pm 1.5$~\cite{Leutwyler1}. There is a relation between the bare quark mass and the pseudoscalar meson mass following from an assumption of PCAC relation, $m_{PS}^{2}  =  C(m_{q_{1}} + m_{q_{2}})$. We determine the coefficient $C$ by fitting a linear function to our data. We then fix the scale $a^{-1}$ for the estimation of the quark masses from a plane of $[(af_{P}), (am_{PS})^{2}]$, using the experimental results of the Kaon decay constant $f_{K^{-}}^{Exp.} = (156.2 \pm 0.2 \pm 0.6 \pm 0.3)$ [MeV] and Kaon mass $m_{K^{-}}^{Exp.} = 493.677 \pm 0.013$ [MeV] (Particle Data Group) as the input values~\cite{Giusti5,Allton1}. We do not consider the errors of the experimental results in our computations. The scale that is determined from $f_{K} (m_{K})$ is $a^{-1} =  2.00(8) \ [\mbox{GeV}]$. We use this scale to evaluate the quark masses. The quark masses are computed from $m_{K}$ and the coefficient $C$ as follows: $m_{s} = \frac{m_{K}^{2}}{C\left(1+\frac{1}{R}\right)}$, $\bar{m} =  \frac{m_{s}}{R}$. Finally, the renormalization group invariant quark mass is estimated as follows: $m^{\overline{MS}} = \hat{Z}_{M}\cdot m$, $\hat{Z}_{M} = z/Z_{S}$, z = 0.72076. The results of quark masses are in Table~\ref{tb:condensate_quarkmass_1}. The quark masses that are computed from normal configurations are compatible with the experimental results. The masses slightly become heavy with increasing the monopole charges as shown in Fig.~\ref{fig:quark_mass_1}. 

\section{Summary and conclusion}
We have confirmed that we properly compute the topological susceptibility and the number of instantons. We have shown that the additional monopoles create instantons. The finite lattice volume and the discretization do not affect the result. The chiral condensate that are computed from the Gell-Mann-Oakes-Renner formula and the random matrix theory are consistent with each other. The quark masses that are computed from normal configurations are consistent with the experimental results. Lastly, the preliminary results show that the chiral condensate decreases and the quark masses become slightly heavy, by increasing the number of monopole charges.

\section{Acknowledgments}
The simulations have been performed on, SX-ACE, SX-9, SX-8, and PC clusters at the Research Center for Nuclear Physics and the Cybermedia Center at the University of Osaka, and SR16000 at the Yukawa Institute for Theoretical Physics at the University of Kyoto. We really appreciate their technical supports and the computation time.

\end{document}